\documentclass{aastex631}

\usepackage{appendix}
\usepackage{booktabs}

\received{March 6, 2023}
\accepted{May 28, 2024}

\submitjournal{ApJ}

\shortauthors{Clocchiatti et al.}

\begin{document}

\title{Global Anisotropies of $\Omega_\Lambda$}

\author[0000-0003-3068-4258]{Alejandro Clocchiatti}
\affiliation{Pontificia Universidad Cat\'olica de Chile \\
Vicu\~na Mackenna 4860 \\
Macul, Santiago, Chile}
\affiliation{Instituto Milenio de Astrof\'{\i}sica (MAS) \\
Nuncio Monse\~nor S\'otero Sanz 100, Of. 104\\
Santiago, Chile}

\author[0000-0001-8651-8772]{\'Osmar Rodr\'iguez}
\affiliation{Instituto Milenio de Astrof\'{\i}sica (MAS) \\
Nuncio Monse\~nor S\'otero Sanz 100, Of. 104\\
Santiago, Chile}

\author[0009-0005-4244-8588]{Ariel \'Ordenes Morales}
\affiliation{Pontificia Universidad Cat\'olica de Chile \\
Vicu\~na Mackenna 4860 \\
Macul, Santiago, Chile}
\affiliation{Instituto Milenio de Astrof\'{\i}sica (MAS) \\
Nuncio Monse\~nor S\'otero Sanz 100, Of. 104\\
Santiago, Chile}

\author[0009-0004-5095-6606]
{Benjamin Cuevas-Tapia}
\affiliation{Pontificia Universidad Cat\'olica de Chile \\
Vicu\~na Mackenna 4860 \\
Macul, Santiago, Chile}
\affiliation{Instituto Milenio de Astrof\'{\i}sica (MAS) \\
Nuncio Monse\~nor S\'otero Sanz 100, Of. 104\\
Santiago, Chile}




\begin{abstract}

An analysis of the Cosmological Constant $\Omega_\Lambda$ fitted to subsamples of the Pantheon+ Type Ia SN sample spanning 2$\pi$ sterradians for a grid of 432 pole positions covering the whole sky reveals two large scale asymmetries.
One of them is closely aligned with the Galactic North-South direction and the other points approximately towards RA$\sim 217.5^\circ$ Dec$\sim -26.4^\circ$, $\sim$50.9 degrees from the CMB dipole Apex.
The signal to noise ratio (S/N) of the multiple $\Omega_\Lambda$ measurements in these directions is $3.2\lesssim $\, S/N\, $\lesssim 8.4$.
The first asymmetry is puzzling, and would indicate a systematic effect related with the distribution of Pantheon+ SNe on the sky and, probably, how the correction for reddening in the Galaxy is calculated.
The second one, which entails a 2.8-$\sigma$ tension between $\Omega_\Lambda$ measure in opposite directions, bears strong implications on its interpretation as Dark Energy: It is consistent with the prediction for tilted observers located in a Friedmann-Robertson-Walker universe who could measure an acceleration or a deceleration with a dipolar asymmetry, irrespective of what the universe as a whole is doing.
In this case, $\Omega_\Lambda$ would not be a physical entity, a real Dark Energy, but an apparent effect associated with the relativistic frame of reference transformation.

\end{abstract}



\section{Introduction} \label{sec:intro}

The effect of proper motions of observers of the cosmos and the kinematics of the local, nearby universe around them, has been essentially disregarded since the pioneer days of supernova (SN) Cosmology     \protect\citep{1995ApJ...440L..41P, 1998ApJ...507...46S}.
They may, however, be playing critical role at determining the nature of the luminosity distances and redshifts we measure, and the subsequent interpretation we make of the resulting Hubble Diagrams.

The papers reporting the discovery of $\Omega_\Lambda$ implicitly assumed that observers on Earth were just sitting on the rest frame of the Universe     \protect\citep{1998AJ....116.1009R,1999ApJ...517..565P} and moved relative to any other with the relative velocity given by the Hubble law.
Care to specify the reference frame of the redshifts, for example, began more than a decade after the discovery of $\Omega_\Lambda$     \protect\citep{2011ApJS..192....1C} as one more ingredient to reduce the, expected small, systematic effects that biased the fitted cosmic parameters.

A different and serious general questioning of these kind of simplifying assumptions which are the core of the, now, traditional SN Cosmology analysis began at about the same time. 
 \protect\cite{2008GReGr..40..269S}, showing that alternative (though contrived) cosmologies could also fit the data, suggests that $\Omega_\Lambda$ ``may well be an artifact of an oversimplified cosmological model'' forced to fit a still not completely understood data set.

Questioning our position as impartial observers tended to align with both observational and theoretical standpoints.
On the observational side many studies have shown that a sizable region around us is taking part of a global motion, a ``bulk-flow'', that has generally not been appropriately accounted for when computing distances and redshifts of the SNe we use to build the Hubble Diagrams.
For a recent and comprehensive review of the relevant observational studies of peculiar velocities and the implied large scale flows in the nearby universe, the reader is referred to  \protect\citet{2022MNRAS.513.2394A}.
On the theoretical side, consideration of contrived alternatives to the standard Friedmann-Robertson-Walker (FRW) background were upgraded by the development of an actual fully relativistic alternative that considers the differences it makes for observers located in a FRW background to be, or not to be, measuring the universe from within a large scale bulk-flow.
 \protect\citet{2011PhRvD..84f3503T, 2021EPJC...81..753T} shows that observers partaking of large-scale, global motions (``tilted'') with respect to the CMB rest frame reside inside volumes where the local kinematics is dominated by peculiar-velocity perturbations rather than by the background Hubble expansion.
This is caused by a surprising relativistic effect, resulting even from non-relativistic peculiar velocities, which triggers a 4-acceleration that makes observers' world-lines no longer geodesics.
The characteristic length scale of these volumes would be between a few and several hundred Mpc depending on the size and velocity of the bulk flows.
The relevant issue is that, within these volumes,  observers may measure an {\em apparent} cosmic acceleration even if the Universe was decelerating.
The seriousness of this theoretical alternative is illustrated by  \protect\citet{2022MNRAS.513.2394A}, who fit a streamlined, parameterized, version of the model to the Pantheon SN sample  \protect\citep{2018ApJ...859..101S} and show that, according to some statistical criteria, the model fits the data as well as the standard $\Lambda$CDM cosmology with the usual assumptions.

Two fruitful main predictions of the new theoretical paradigm have been highlighted by  \protect\citet{2022EPJC...82..521T}.
One is that ``tilted'' observers will measure a change of sign of the deceleration parameter as if the universe had switched from deceleration to acceleration rather recently, as it has been actually observed  \protect\citep[e.g.][]{2007ApJ...659...98R}, the other is that they will detect a dipolar anisotropy of the deceleration parameter if they measure its full-sky distribution.
An additional prediction is that the dipole of the deceleration/acceleration parameter should be oriented closely to the CMB dipole since both would originate in the same local rest frame motion.

In this paper we look for large scale anisotropies in $\Omega_{\Lambda}$ using the Pantheon+ SN sample  \protect\citep{2022ApJ...938..110B} and we show that they are actually present.
In section \ref{se:design} we describe our strategy for the analysis and present the results.
In section \ref{se:discussion} we discuss these results, try to assess the significance of our findings by analyzing with the same method two different kinds of random variations of the Pantheon+ sample, and fit dipoles to the detected signals.
Finally, in section \ref{se:final} we summarize the results and give our conclusions.

\section{Set up of the analysis} \label{se:design}
\subsection{Data} \label{ss:data}
\begin{figure}[ht!]
\plottwo{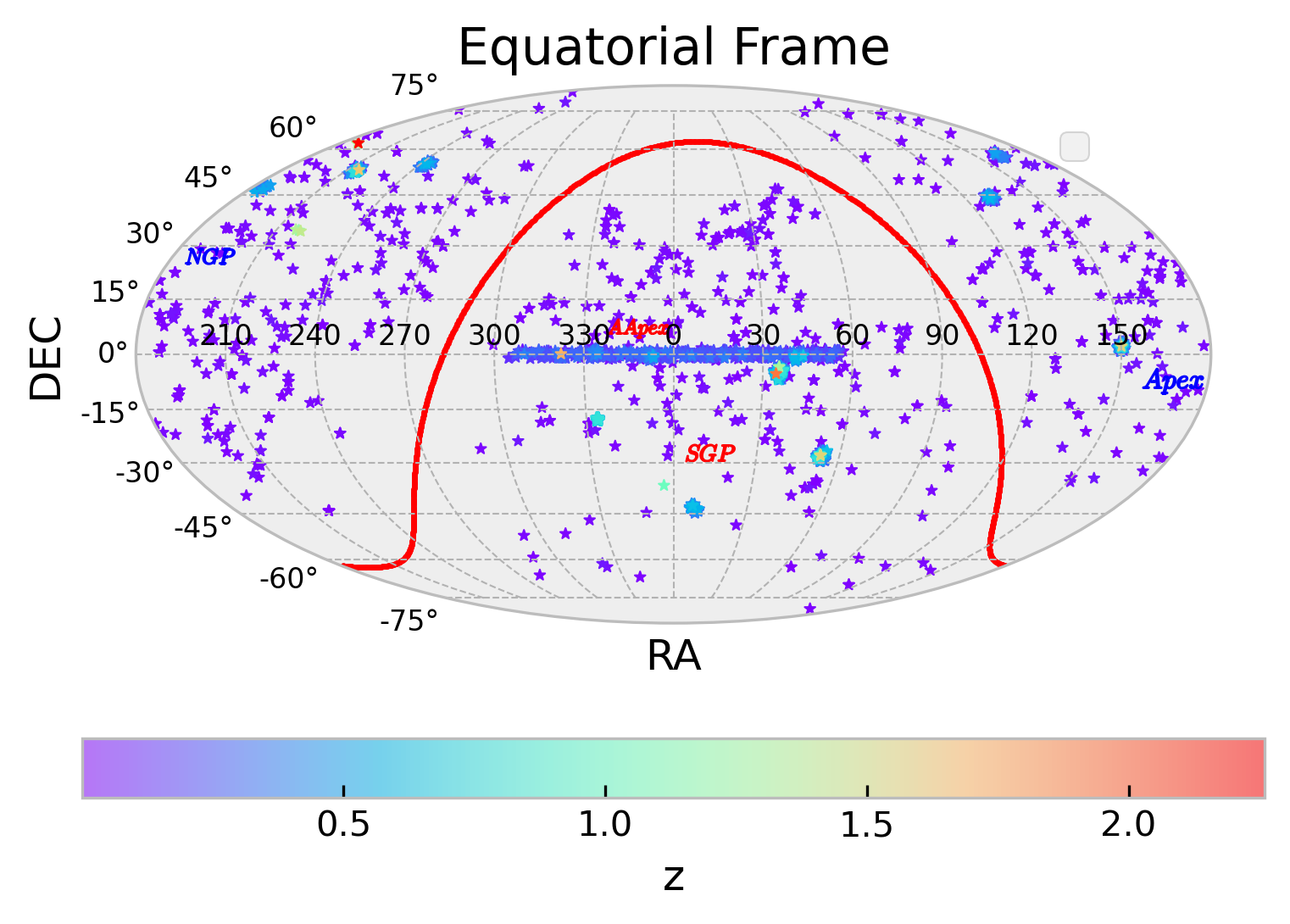}{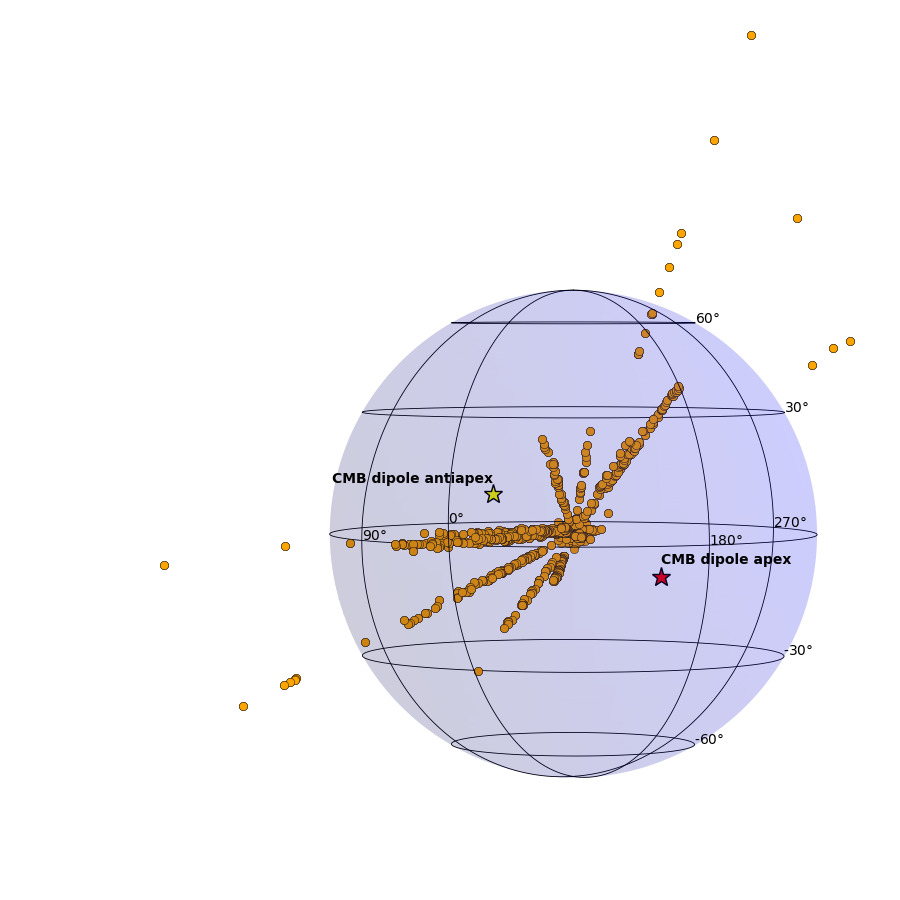}
\caption{Left panel: Distribution of the SNe in the Pantheon+ sample on the sky in a Mollweide projection, color coded by redshift. The solid red line shows the Galactic Equator and ``NGP'' and ``SGP'' mark the positions of the North and South Galactic Poles, respectively. Apex and AApex labels mark the positions of the CMB Apex and Anti apex, respectively.
Right panel: Projected 3D distribution of the Pantheon+ SN sample. The third dimension is the redshift $z$, and the sphere represents $z = 1$. The kinematic CMB Apex and Antiapex positions have been marked by stars. The Apex side is towards the reader.}
\label{fi:sample_in_sky}
\end{figure}

The Pantheon+ SN sample  \protect\citep{2022ApJ...938..110B} is a compilation of 1701 light curves, of 1550 cosmology grade Type Ia SNe with redshifts between $z\sim 10^{-3}$ and $z\sim 2.3$.
The distribution of redshifts is fairly non-uniform {(see e.g. Figure~1 of   \protect\citealt{2022ApJ...938..113S})}, with one third of the sample below $z\sim 0.04$ and two thirds of the sample below $z\sim 0.27$.
The uncertainties in the distances are non-uniform, as well. They are larger at redshifts smaller than $z\sim 0.1$, minimal between $z\sim 0.1$ and $z\sim 0.3$ and steadily growing with a moderate tendency beyond that.
The uncertainties in the redshift are more uniform, with $\sim 95$\% of the SNe having uncertainties below 0.01 ($\sim 90$\% below 0.005).
Finally, the distribution of the SNe on the sky is highly inhomogeneous, with many of those at low to intermediate redshifts concentrated on a $\sim 120^{\circ}$ RA equatorial strip centered on RA$= 0 ^{\circ}$ and, naturally, most of them located at high Galactic latitudes (see the left panel of Figure~\ref{fi:sample_in_sky}).
Also relevant is that most of those at high redshifts are concentrated towards the North and South Galactic Poles, and that there is a wide solid angle devoid of high redshift SNe around the direction of the CMB Apex (see the right panel of Figure~\ref{fi:sample_in_sky}).

  \protect\citet{2022ApJ...938..110B} provide three realizations of the redshift towards each Pantheon+ SN:
$z_{\rm He}$ the heliocentric redshift,  $z_{\rm CMB}$, the heliocentric redshift corrected by the motion of the Sun with respect to the Cosmic Microwave Background, and $z_{\rm HD}$, the previous one corrected in addition by the peculiar velocities of the SNe parent galaxies (see their \S 3.1.3 and their Table~7).
We have used in our study $z_{\rm HD}$.
The data was obtained from https://github.com/PantheonPlusSH0ES/DataRelease.

\subsection{Partitioning the sample} \label{ss:partition}

Measuring anisotropies in $\Omega_\Lambda$ requires doing a complete fit of the cosmology on SNe samples located in different regions of the sky.
There are different ways of doing so. If it is true that $\Omega_\Lambda$ displays a dipolar anisotropy caused by a somewhat different composition of the same local motions that cause the CMB dipole   \protect\citep{2022EPJC...82..521T}, the strategy that maximizes the difference to be found between two positions in the sky is to fit the cosmology to a complete SN sample (with a healthy number of both low redshift and high redshift SNe) located in a small solid angle around the direction to the CMB Apex, and compare it with a fit done on a similar SN sample located in a small solid angle towards the CMB Anti Apex. Unfortunately, those samples do not presently exist.
Splitting Pantheon+ in subsamples located in wider ``cusps'' and rings of varying solid angles around the CMB Apex and Anti Apex generally provides subsamples with very different histograms of redshifts, distances, and associated uncertainties.
Since the $\chi^2$ minimization used to fit the cosmological parameters depends critically on the redshift and uncertainty distribution of the SN sample, those differences imply a strong bearing on the fitted cosmological parameters, which complicates the subsequent comparison of the results.
The ideal of having SN samples with many hundreds of cosmology grade Type Ia SNe well distributed in redshift in multiple small patches of the sky conveniently located for this analysis will have to wait until surveys like the Vera C. Rubin LSST   \protect\citep{2022ApJS..258....1B} are advanced.

A compromise solution is to analyze the currently available samples using the hemisphere comparison method   \protect\citep{2007A&A...474..717S,2010JCAP...12..012A} since by maximizing the ``cusp'' around each of the chosen directions, the strategy provides SN samples with more similar parameter distribution for both hemispheres.


We selected a set of directions on the celestial sphere using the tools of the {\em Hierarchical Equal Area and iso-Latitude Pixelation}   \protect\citep[HEALPix\footnote{https://healpix.sourceforge.io},][]{2005ApJ...622..759G} package, choosing a grid resolution parameter $N_{side} = 6$, which provides 432 equal area pixels on the sky.
This gives us 216 axes, each one with an ``Up'' and a ``Down'' pole.
Each Up and Down pole pair defines a Northern-like hemisphere and a Southern-like hemisphere, and determines the SNe subsamples that fall into each of them.

In a different version of this experiment we selected the directions on the celestial sphere using a Fibonacci Lattice   \protect\citep{Gonz_lez_2009} with 420 points (210 axes).
The results of this selection are qualitatively similar to the first one and will not be presented here, although some mention of the differences will be made later on. 

In assigning SNe to an Up or Down hemisphere, an additional issue to consider is the local subsample.
The effect of local motions is stronger on the redshifts of nearby SNe, which are the anchor of the Hubble diagrams   \protect\citep{2019MNRAS.490.2948D}.
But, on the other hand, we have better prospects of correcting for the local motions those SNe which are closer to us.
In order to give our analysis some resilience against systematic effects associated with the local sample we first followed   \protect\citet{2022ApJ...934L...7R} and   \protect\citet{2022ApJ...938..110B} and rejected those SNe with redshifts below $z = 0.01$, if the distances of their host galaxies were not measured with Cepheids, and, second, we assumed that redshifts below $z = 0.02$ are correctly transformed to the CMB frame and provide a reliable anchor for the Hubble Diagrams.
The two criteria leave 221 Pantheon+ SNe below $z = 0.02$, a local sample, which were included in all the Hubble diagrams, irrespective of the chosen pole direction (i.e. our hemispherical subsamples of SNe are different only for redshifts larger than 0.02).

Another issue to discuss  is that of the systematic uncertainties of, and the covariances between, Pantheon+ SNe. In this work we are not trying to reach a definitive, unbiased, value of the cosmological parameters, but analyzing the relevance of the location of subsamples on the celestial sphere, and hence modifying the subsamples as the Up and Down poles considered shift their position on the sky.
As we will describe later, one of the tests we did involved reshuffling the locations of the SNe on the celestial sphere.
This modifies some of the assumptions on which both the systematic uncertainties and the covariance matrices are built.
Hence, the covariance matrices provided by the Pantheon+ collaboration are not directly applicable to our hemispheric analysis, so it is preferable for 
this analysis to assume (1) that the uncertainties in SN distances are uncorrelated, and (2) restrict the analysis to the statistical uncertainties.
This will allow us to produce results that are internally comparable, though they will be systematically biased.
As we show in the Appendix, however, the differences in the fitted cosmological parameters implied by these assumptions are very minor in comparison with the differences that result from grouping the SNe in opposite hemispheres and, then, will not interfere with interpreting our results. 

\subsection{Results} \label{ss:results}

Our experiment consisted, hence, in choosing 216 Up poles in the sky, selecting the Pantheon+ SNe that fall within 2$\pi$ sterradians of each one and performing a full $\chi^2$ fit of the cosmological parameters for these subsamples of SNe as well as for the complementary subsample within 2$\pi$ sterradians of the corresponding Down pole.
The fit to the 432 different Hubble Diagram was done using the theoretical expression for the luminosity distance at a redshift $z$ in a FRW cosmology 
  \protect\citep[e.g. eq.~2 in][]{1998AJ....116.1009R} for a fine grid of $\Omega_\Lambda$, $\Omega_M$ and $H_{\rm 0}$, marginalizing over the latter, and finding the minimum of $\chi^2$ in the $\Omega_\Lambda$ and $\Omega_M$ space.
This provides the joint confidence intervals, the best fit values of $\Omega_\Lambda$ and $\Omega_M$ and, from the marginal distributions, an estimate of their individual uncertainties.
We emphasize that the described procedure makes the fitted $\Omega_\Lambda$ and $\Omega_M$ pair independent of the actual value of $H_{\rm 0}$ (and, hence, of the actual Type Ia SN intrinsic luminosity), and naturally accommodates universes with non-zero curvature.

The result of the hemispheric analysis are presented in Figure~\ref{fi:mollweide_real} as a color map of the value of the $\Omega_\Lambda$ parameter fitted in each of the 432 selected positions on the sky, in Figure~\ref{fi:mollweide_diff_real} as a color map of the difference between the value of $\Omega_\Lambda$ computed for the Up and Down poles, and in Figure~\ref{fi:OM_high_low} as the confidence contours in the $\Omega_\Lambda$ and $\Omega_M$ plane for the direction that provides the largest difference between the Up and Down poles.
In both Figures~\ref{fi:mollweide_real} and \ref{fi:mollweide_diff_real} the right panel plots the signal to noise ratio of the measurements given in the left panel.

\subsubsection{Sanity check}
We checked that the results of   \protect\citet{2022ApJ...938..110B} for the full all-sky Pantheon+ SN Sample are recovered by our analysis by applying the same method to the full Pantheon+ sample.
Specifically, we tested the difference it makes in the cosmological parameters to use the whole covariance matrix   \protect\citep[i.e. using eq. 15 in ][]{2022ApJ...938..110B} or assume that the uncertainties are uncorrelated    \protect\citep[i.e. using eq.~2 in][]{1998AJ....116.1009R}.
We found  that the differences do exist, but do not alter significantly the results reported here. 
Further details on the  comparison  of the two methods are given in the Appendix.

\subsubsection{Variation of $\Omega_\Lambda$ with direction in the sky}

\begin{figure}[ht!]
\plottwo{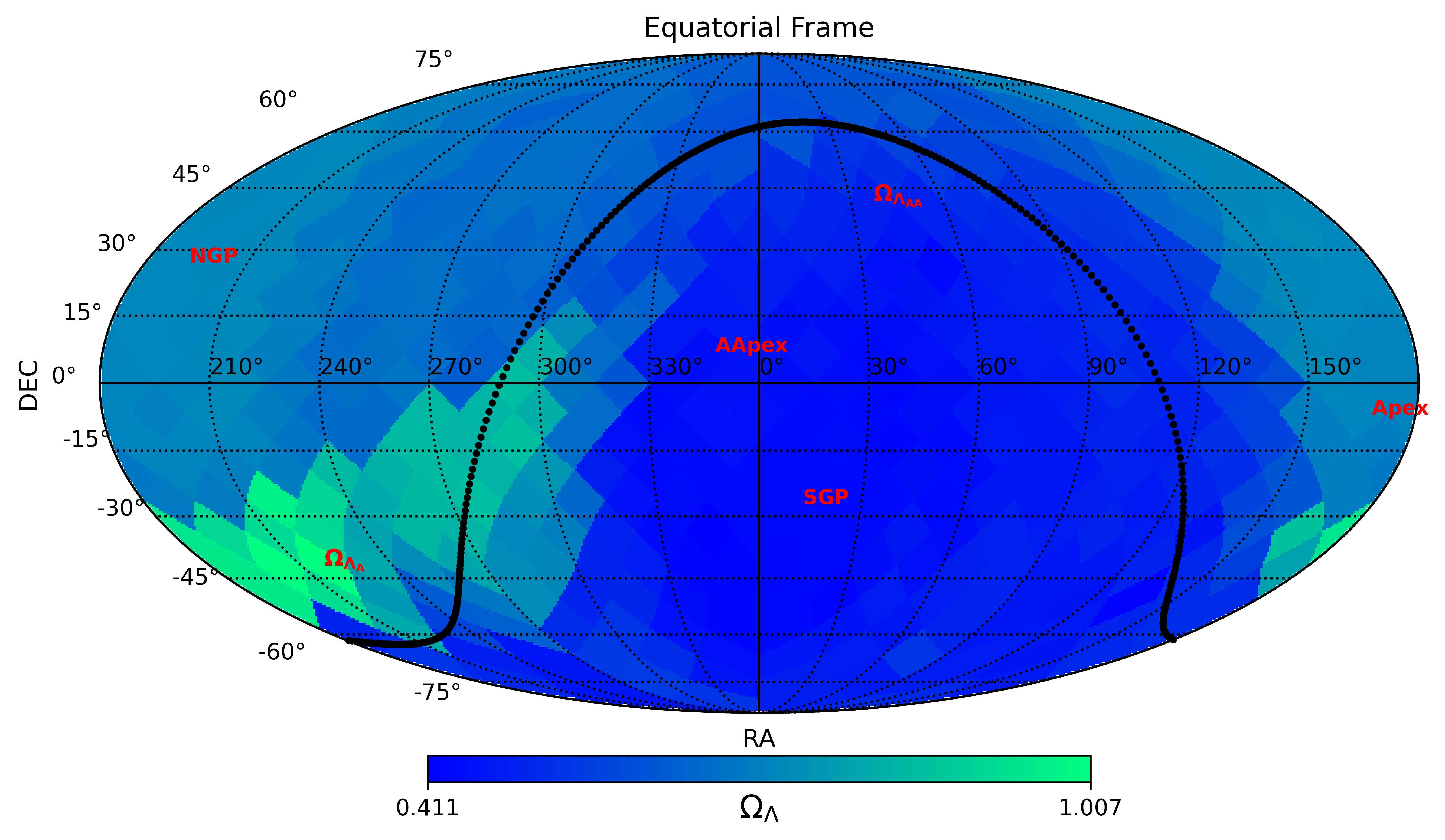}{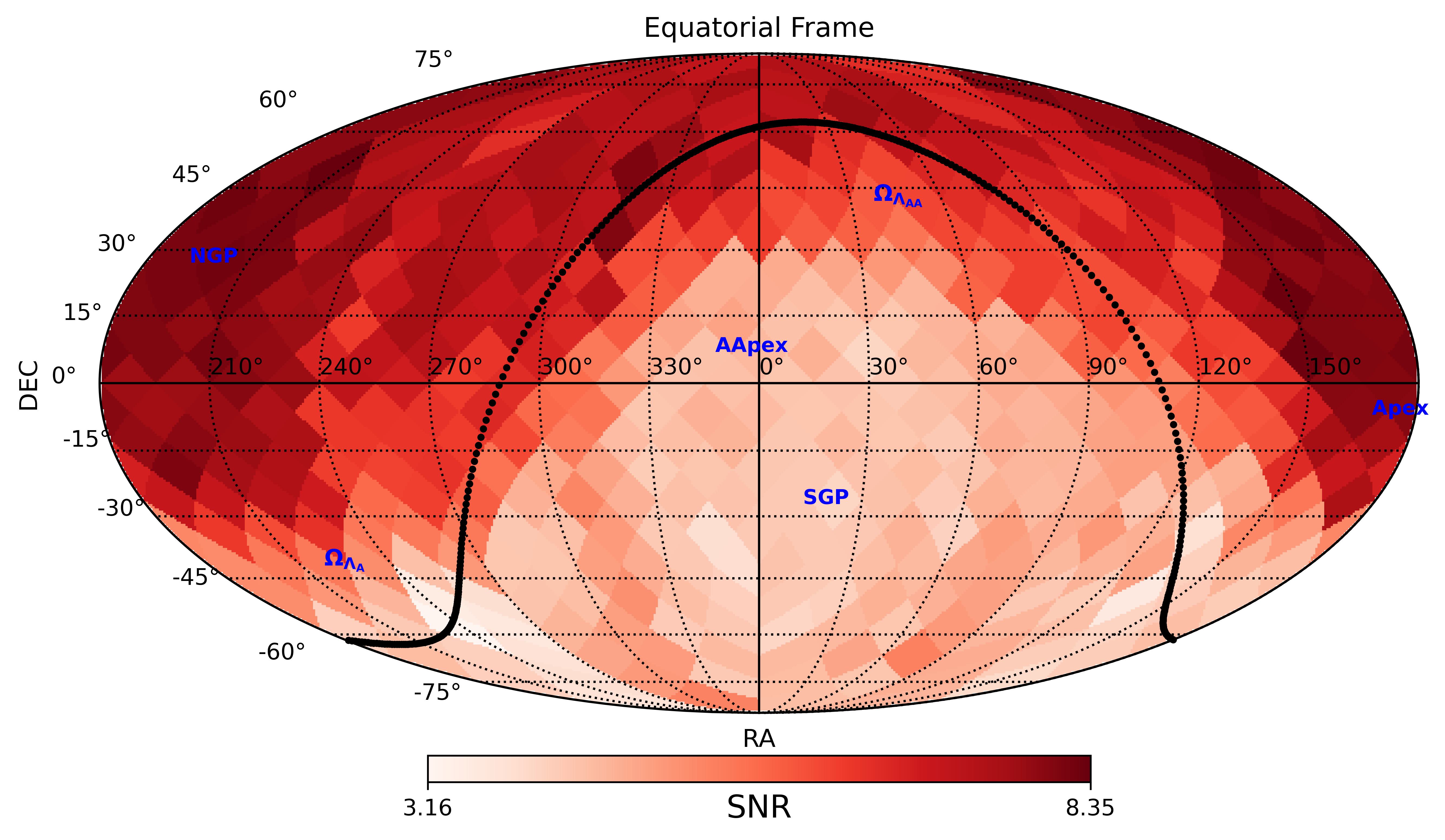}
\caption{Color map of the variation of $\Omega_\Lambda$ over the whole sky, shown in Mollweide projection using Celestial Equatorial Coordinates (left panel). The solid black line shows the Galactic Equator and ``NGP'' and ``SGP'' mark the positions of the North and South Galactic Poles, respectively. Labels Apex and AApex mark the positions of the CMB Apex and Anti apex, respectively. 
$\Omega_{\Lambda A}$ and $\Omega_{\Lambda AA}$ mark the positions were the fitted values of $\Omega_\Lambda$ Apex and Antiapex, respectively, provide the largest difference (see text). 
The right panel shows the signal to noise ratio of these measurements.}
\label{fi:mollweide_real}
\end{figure}

The results plotted in Figure~\ref{fi:mollweide_real} show that $\Omega_\Lambda$ varies across the sky, and the variation displays two large scale patterns.
There is a large collection of low values surrounding the South Galactic Pole and a corresponding large gathering of high values around the North Galactic Pole.
These two areas compound the asymmetry that covers the largest surface on the sky.
There is, in addition, a clear patch of very large values of $\Omega_\Lambda$ that extends from RA$\sim 300^\circ$ Dec$\sim 0^\circ$ towards RA$\sim 170^\circ$ Dec$\sim -40^\circ$. 
This area is smaller than the previous two ones, but includes the highest values, with $\Omega_\Lambda \sim 1$. We will call this patch the $\Omega_\Lambda$ Apex region.

The smallest value of $\Omega_\Lambda$ is found at RA$\sim 345^\circ$ Dec$\sim -34^\circ$ ($\Omega_\Lambda = 0.41 \pm 0.11$).
The signal to noise ratio of these measurements (SNR), estimated as the ratio between $\Omega_\Lambda$ and its 1-$\sigma$ uncertainty, is greater than three in all cases, as can be seen in the right panel of the figure.

\begin{figure}[ht!]
    \plottwo{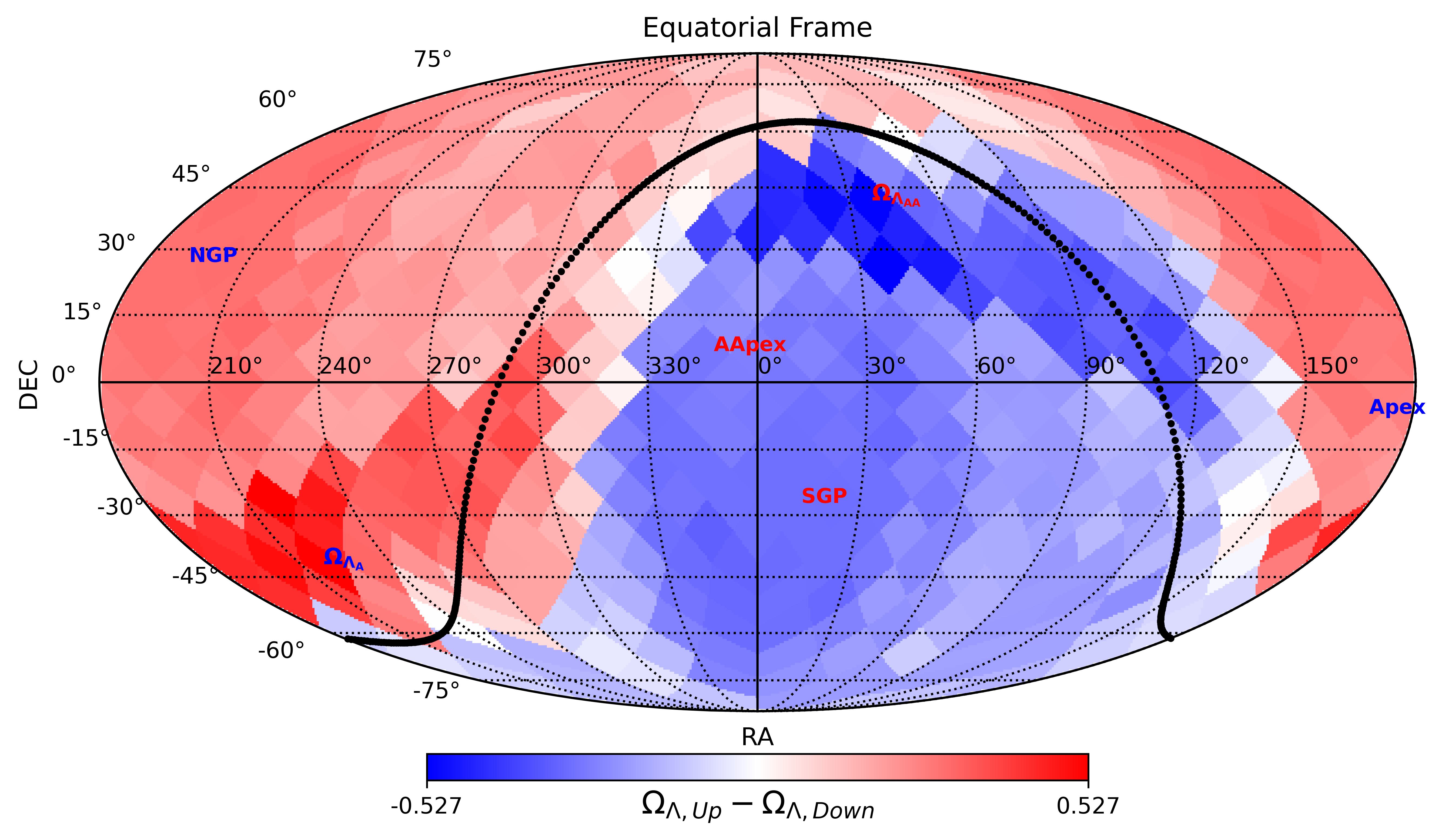}{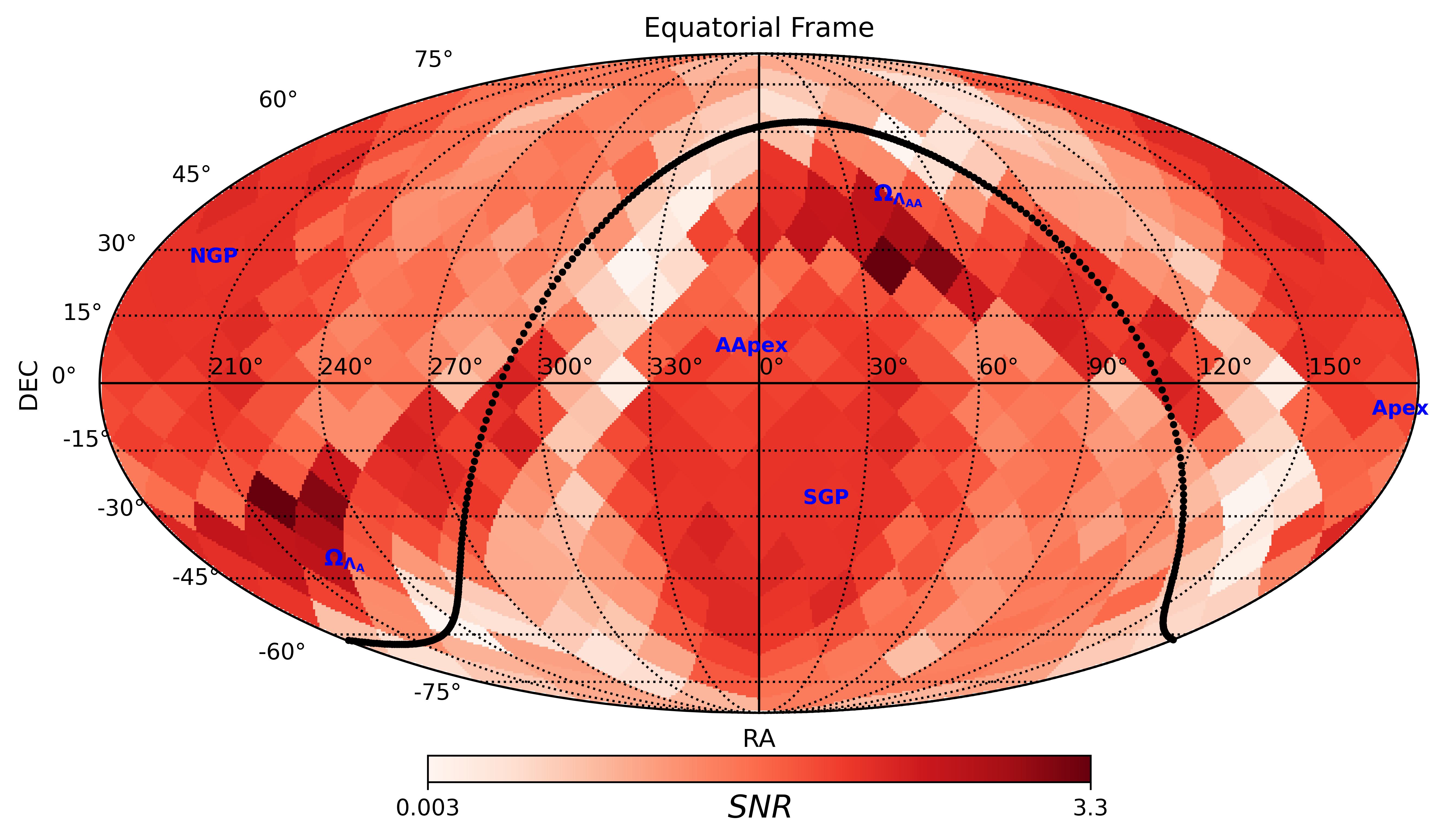}
    \caption{The difference between   $\Omega_\Lambda$ in the Up-Down directions for the 216 axes used in our study (left panel). The right panel displays the SNR of these differences. The figures show 432 points on the sky, with antipodes having the same difference with opposite sign. }
    \label{fi:mollweide_diff_real}
\end{figure}

In Figure~\ref{fi:mollweide_diff_real} the results are shown as difference between the $\Omega_\Lambda$ values measured in opposite directions on the sky (the pole-antipole, or Up-Down, differences).
We note that in the latter figure the Pole-Antipole points carry just the same difference with opposite sign.

The differences Up-Down generally trace the same pattern as the $\Omega_\Lambda$ measurements, with the highest SNR concentrating in the $\Omega_\Lambda$ Apex region and in the directions of the Galactic poles .
The largest absolute value of a difference is found on the axis from RA$\sim 217.5^\circ$ Dec$\sim -26.4^\circ$, with $\Omega_\Lambda = 0.98 \pm 0.13$, to RA$\sim 37.5^\circ$ Dec$\sim 26.4^\circ$, with $\Omega_\Lambda = 0.45 \pm 0.09$, providing a variation of $\Delta \Omega_\Lambda = 0.53 \pm 0.16$. In what follows, we will call these directions the $\Omega_\Lambda$ Apex and Antiapex, respectively.

The smallest absolute values of the difference trace the transitions between positive and negative values of  $\Delta \Omega_\Lambda$ , which given the large pattern aligned with the Galactic rotation axis tend to be close to the Galactic plane.

{ In figure \ref{fi:OM_high_low} we show the confidence contours of  $\Omega_\Lambda$ and $\Omega_M$  
for the  $\Omega_\Lambda$ Apex--Antiapex axis.
The contours towards the $\Omega_\Lambda$ Apex are plotted in blue and those towards the Antiapex in red.
As a guide to the eye, the confidence contours for the whole Pantheon+ sample are shown in black.
The results on the left panel correspond to the color maps of Figures~\ref{fi:mollweide_real} and \ref{fi:mollweide_diff_real}. They are computed using only the statistical uncertainties and assuming that they are uncorrelated.
As a comparison, we give in the right panel the results computed for the same Apex--Antiapex axis using the complete statistical plus systematic covariance matrix, {$C_{\rm stat+sys}$}   \protect\citep[eq. 8 in][]{2022ApJ...938..110B}. Further detail on this last calculation are given in the Appendix.
\begin{figure}[ht!]
\plottwo{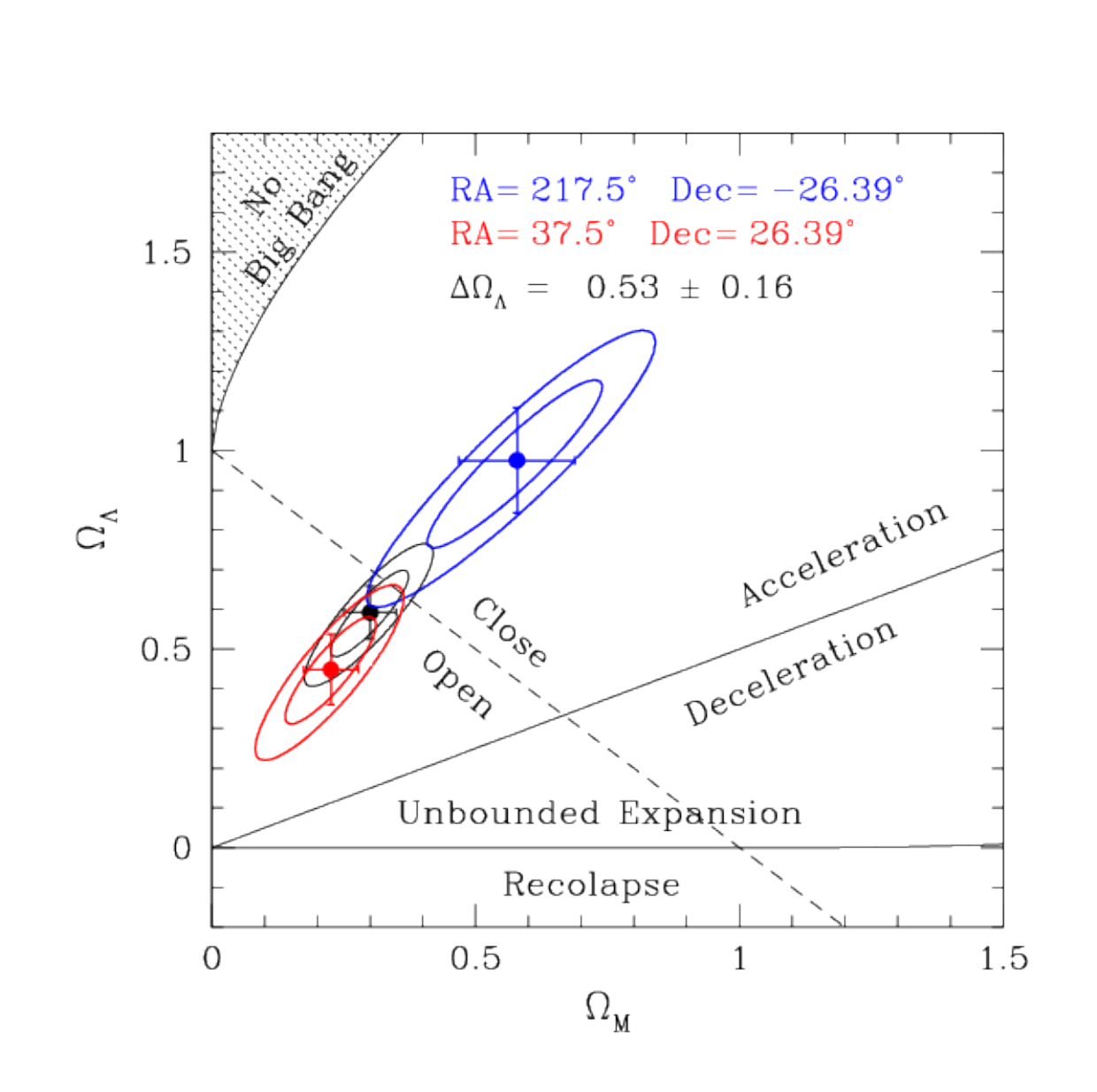}{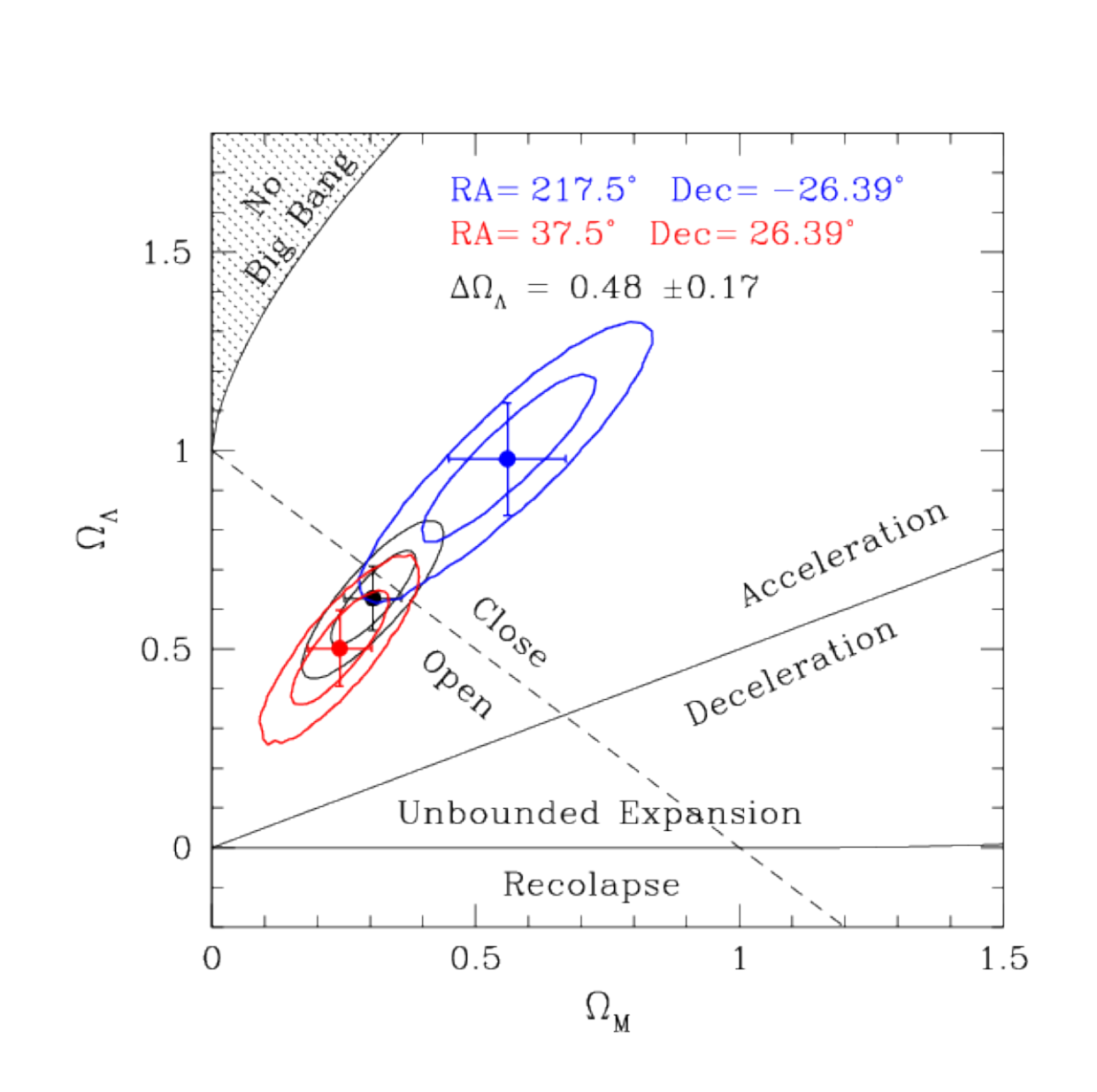}
\caption{
Joint confidence intervals of $\Omega_\Lambda$ and  $\Omega_{\rm M}$ for the full Pantheon+ sample (black lines) and the Up and Down subsamples in the direction that provides the maximum difference in $\Omega_\Lambda$.
Blue lines are used for the contours in the direction of the largest $\Omega_\Lambda$ and red lines for the direction of smallest $\Omega_\Lambda$.
The left panel shows the results of using the diagonal of $C_{\rm stat}$, and the right panel those of using the whole $C_{\rm stat+sys}$ covariance matrix.
In both cases, the contours correspond to the 68.3\%, 95.4\% probability levels.
}
\label{fi:OM_high_low}
\end{figure}
The best fit cosmological parameters, which are detailed in Table~\ref{ta:compApexAapex} of the Appendix, are plotted in Figure~\ref{fi:OM_high_low} as well.
The uncertainties are ``1-$\sigma$'' computed from the one dimensional marginal distributions (i.e they enclose 68.3\% of the marginal probability distribution around the maximum value).
The fitted values for the whole sample
($\Omega_{\rm M}= 0.300 \pm 0.050$, 
$\Omega_\Lambda = 0.592 \pm 0.067$, in case of the left panel, and
$\Omega_{\rm M}= 0.305 \pm 0.054$, 
$\Omega_\Lambda = 0.628 \pm 0.080$, in case of the right one)} are consistent with those of   \protect\citet{2022ApJ...938..110B}, as can be seen in the second line of the first block of their Table~3, where they report $\Omega_{\rm M}= 0.306 \pm 0.057$ and $\Omega_\Lambda = 0.625 \pm 0.084$.
The later values, as well as the contours in Figure~8 of the cited paper, are computed with Pantheon+ SNe and SH0ES distances.
This means a specific calibration of Type Ia SNe intrinsic luminosity and, hence, have a direct implication on the value of $H_0$ reported on column 4 of that table, but have no bearing on the fitted $\Omega_\Lambda$ and $\Omega_{\rm M}$ parameters which are obtained after marginalizing on $H_0$.

\section{Discussion} \label{se:discussion}

The hemispherical analysis of the Pantheon+ SN sample indicates that the Cosmological Constant varies with direction in the sky.
The differences between the fitted $\Omega_\Lambda$ in varying positions span a factor larger than two, from 0.41 to 1.01.
This is reassuring regarding us not using the Pantheon+ covariance matrix to compute $\chi^2$, since its impact on the fitted cosmology is much smaller than that (see Table~\ref{ta:compApexAapex} in the Appendix).
{As we discuss below, however, the different methods imply differences in the statistical significance of the $\Omega_\Lambda$ Apex-Antiapex contrast.}

\subsection{Assessing the significance}

Are the variations of $\Omega_\Lambda$ shown in Figure~\ref{fi:mollweide_real} significant? There are at least two issues to consider.
One is the significance of the individual differences, in particular those in each 
Up--Down hemisphere, which result from independent distant samples.
The other issue is the significance of the global patterns of these variations on the sky.

{ For the cosmological fits shown in the left panel of Figure~\ref{fi:OM_high_low} the Apex-Antiapex axis, between RA$\sim 217.5^\circ$ Dec$\sim -26.4^\circ$ and its antipode, provides a 3.3-$\sigma$ difference.
If, in this same directions, the cosmologies were fitted using the whole $C_{\rm stat}$ covariance matrix the difference would be
$\Delta \Omega_\Lambda = 0.54 \pm 0.15$, a 3.5-$\sigma$ result, and, in the case of the right panel of Figure~\ref{fi:OM_high_low}, when
the whole $C_{\rm stat+sys}$ covariance matrix is used, the difference is a 2.8-$\sigma$ result, with a probability smaller than 0.52\% of being due to chance (see the Appendix for further details).
The fact that the maximum Up-Down contrast shifts from inconsistency (larger that 3-$\sigma$) to tension (larger than 2-$\sigma$) when the systematic uncertainties are included, highlights the prominence of the latter and the need to determine them accurately.}

Another relevant issue is the direction of the Apex-Antiapex axis.
It falls $\sim51$ degrees away from the CMB kinematic Apex. The probability that this angular distance results from chance is 18.5\%.

Taking the results as real, our interest now can be expressed in two questions. (1) How stable are the large asymmetries in $\Omega_\Lambda$ against variations in the sample? and (2) How significant is to find these asymmetries in the Pantheon+ SN collection?

To answer the first question we studied the resilience of the global structures against filtering the SNe with largest uncertainties and variations in the local sample (i.e. the nearby SN sample that all the poles have in common), and performed a bootstrap study of the distant sample.

In the first experiment we removed from the original sample the 36 SNe with uncertainty in the distance modulus larger than 0.5 mag and the 2 with uncertainty in the redshift larger than 0.03, and then applied the same criteria as before to select subsamples and repeated the hemispherical analysis of $\Omega_\Lambda$ over the whole sky.
The color map of the results with this reduced sample is essentially the same one shown in the left panel of Fig.~\ref{fi:mollweide_real}.
All the differences between computed $\Omega_\Lambda$s at the same poles are fully consistent with no difference.

The scan of the sky that produced the results in Fig.~\ref{fi:mollweide_real} was done assuming a common local sample (i.e. the SNe always present in the Hubble diagram irrespective of the chosen hemispheric direction) composed of those SNe with $0.01 \le z \le 0.02$, and those at smaller redshift which are Cepheid calibrated.
In the second experiment we tested the influence of this choice by repeating the whole computation with local samples made up of SNe with $0.01 \le z \le 0.04$ and $0.01 \le z \le 0.06$ (plus the nearer SNe with Cepheid distances).
In both cases, we found basically the same maps shown in Figures~\ref{fi:mollweide_real} and \ref{fi:mollweide_diff_real}, with the values of  $\Omega_\Lambda$ computed with different local samples at the same position on the sky being consistent within 1-$\sigma$ of each other.

A final test related to question one was to bootstrap the sample.
We kept the local sample fixed but retrieved random samples, with reposition, of the Pantheon+ distant SNe subsample up to complete the total distance modulus--redshift pairs of the original sample.
We did a sequence of 100 bootstraps and checked the resulting cosmologies with the same hemispheric analysis used for the real sample.
We found, again, that the structures displayed in Figures~\ref{fi:mollweide_real} and \ref{fi:mollweide_diff_real} are very stable.
The uncertainty of the average $\Omega_\Lambda$ computed in each of the 432 directions in the sky resulted about ten times smaller than the uncertainty from the marginal distribution of each $\Omega_\Lambda$ in the ($\Omega_\Lambda$, $\Omega_{\rm M}$) plane.

Question two, regarding the significance of the detected global structures, focus on the possibility that the particular set of positions, distances, redshifts and uncertainties in the Pantheon+ sample, combined with the smearing effect of the hemispheric sweeping and comparison, could naturally produce this kind of results irrespective of their physical reality.
Since the observed structures strongly depend on the positions of the SNe on the sky, we devised an experiment consisting in shuffling at random those positions.
This is, we rearranged at random the pairs (RA, Dec) with the pairs (redshift, distance modulus) and their uncertainties.
By doing so, the Hubble diagram of the complete Pantheon+ sample is always preserved, together with the global fit to cosmology, but the subsamples in each direction are different.
We repeated this experiment a hundred times and for each new realization of the sample we performed the hemispherical analysis scanning the whole sky with the same 432 positions as before and found the best fitting cosmology in each case.
We then studied the statistics of $\Omega_\Lambda$ fitted values, those of the Up-Down (pole-antipole) differences, and how well a dipole can be fitted to each shuffling.
By scrambling the SN positions at random we altered some of the basis on which the Pantheon+ systematic uncertainties and covariance matrices had been calculated, which is one of the reasons why we used just the statistical uncertainties given by the diagonal matrix elements.

\begin{figure}[ht!]
\plottwo{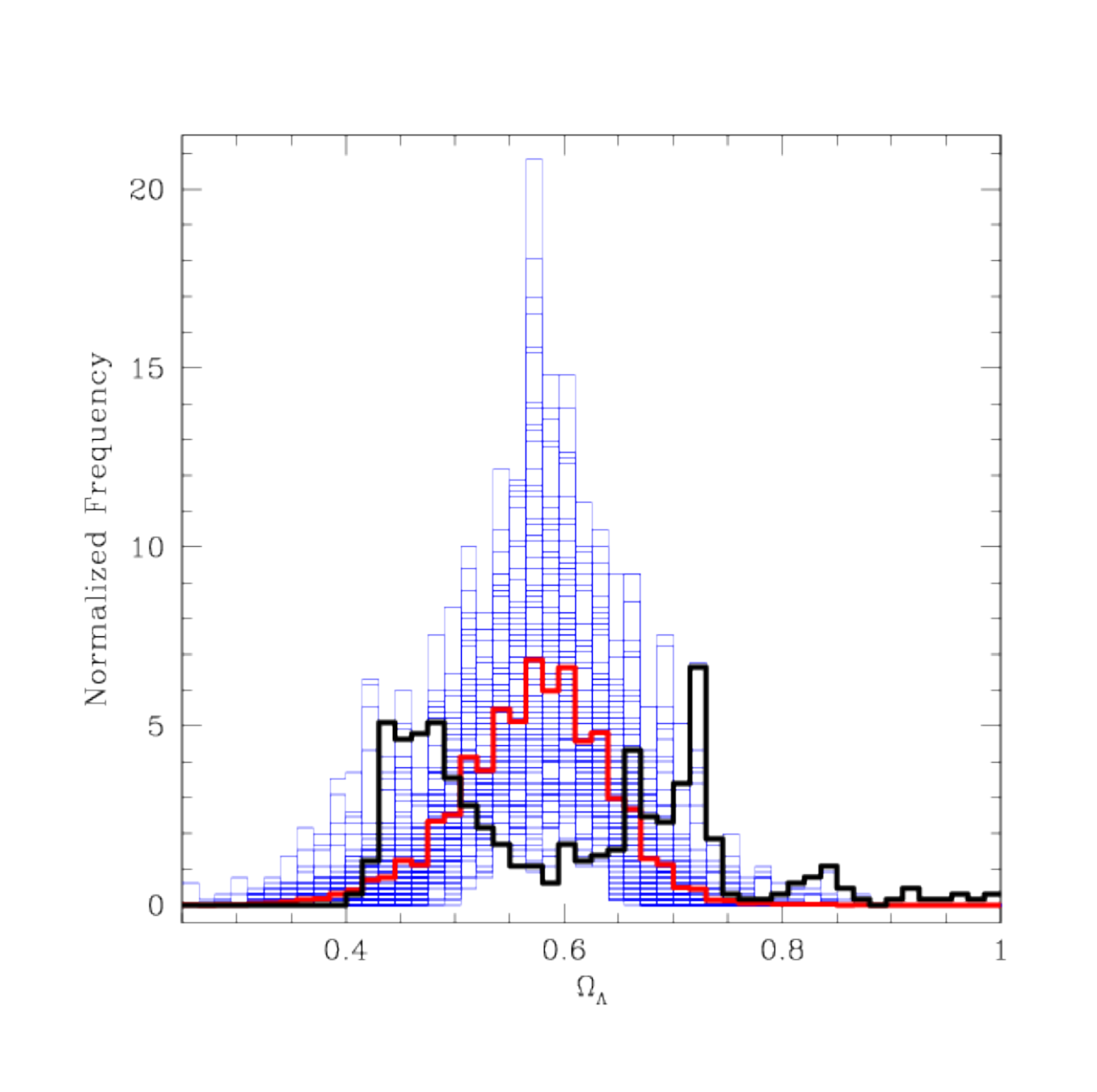}{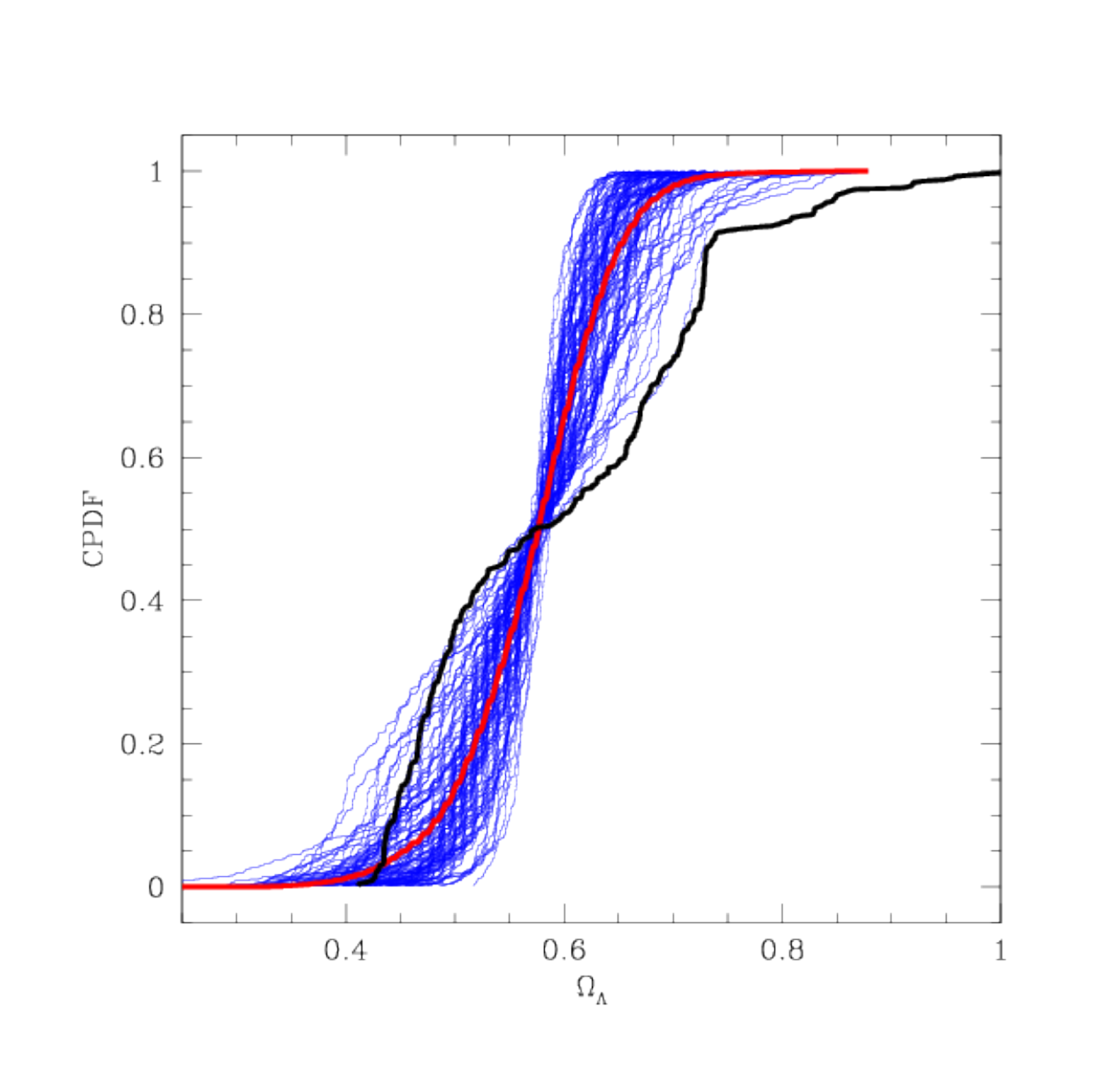}
\caption{
Histograms of the fitted $\Omega_\Lambda$, normalized to unit area, for the 100 realizations of the universe resulting from randomly exchanging the positions of Pantheon+ SNe in the sky (left panel).
Histograms of each individual swap are plotted in light blue lines.
The joint histogram of all 100 realizations is given in thick solid red line.
The histogram of the real universe (the plain Pantheon+ sample) is given in thick black line.
The right panel shows the same statistics through the Cumulative Probability density functions.
The color code is the same as in the left panel.
}
\label{fi:OML_stats}
\end{figure}

In the left panel of Figure~\ref{fi:OML_stats} we present the histograms of the $\Omega_\Lambda$ values measured for each fake distribution of SNe in the sky and in the right panel their Cumulative Probability Density Functions (CPDF), in each case compared with those of the real universe.
The histograms are normalized to a unit area.
It is possible to see that the fake SNe distributions tend to a histogram centered at $\Omega_\Lambda = 0.572 \pm 0.068$,
which is one more indication that SNe in the Pantheon+ sample do measure a cosmological constant with high certainty. Irrespective of how they are arranged in the sky, a cosmological constant is always fitted with histogram mean values between 0.547 and 0.596. The histogram of the real universe has a mean value of 0.597.
The smallest hemispherical value fitted in one of the 43,200 realizations is $\Omega_\Lambda = 0.202 \pm 0.115$.
What you would expect from a $\Omega_\Lambda$ that is really constant with direction in the sky is a histogram like the one plotted with red line, where most axes provide a small difference with the mean value and the large differences are less frequent.
But, although there are a few realizations of the Pantheon+ sample that give broad, flat, or bimodal histograms, that of the real universe (plotted in thick black line) is the most discrepant: It is clearly bimodal and boasts a unique concentration of large values of acceleration, which are associated with the $\Omega_\Lambda$ Apex region.
 
The right panel of the figure depicts a similar view.
Most of the realizations of the sample define what would be a reasonable mean CPDF for a standard shuffling of positions.
There are a few exceptions but, again, the real distribution of SNe in the Pantheon+ sample is remarkably discrepant.
The tail of high values starting at $\Omega_\Lambda \sim 0.7$ is particularly noticeable.
Comparing the histogram Standard Deviations (STDEVs) complements that view: the real distribution of SNe carries the largest STDEV, $\sigma_{\rm Pantheon+}= 0.132$ and 91 out of the 100 fake distributions have STDEV smaller than 0.7$\sigma_{\rm Pantheon+}$.

\begin{figure}[ht!]
\plottwo{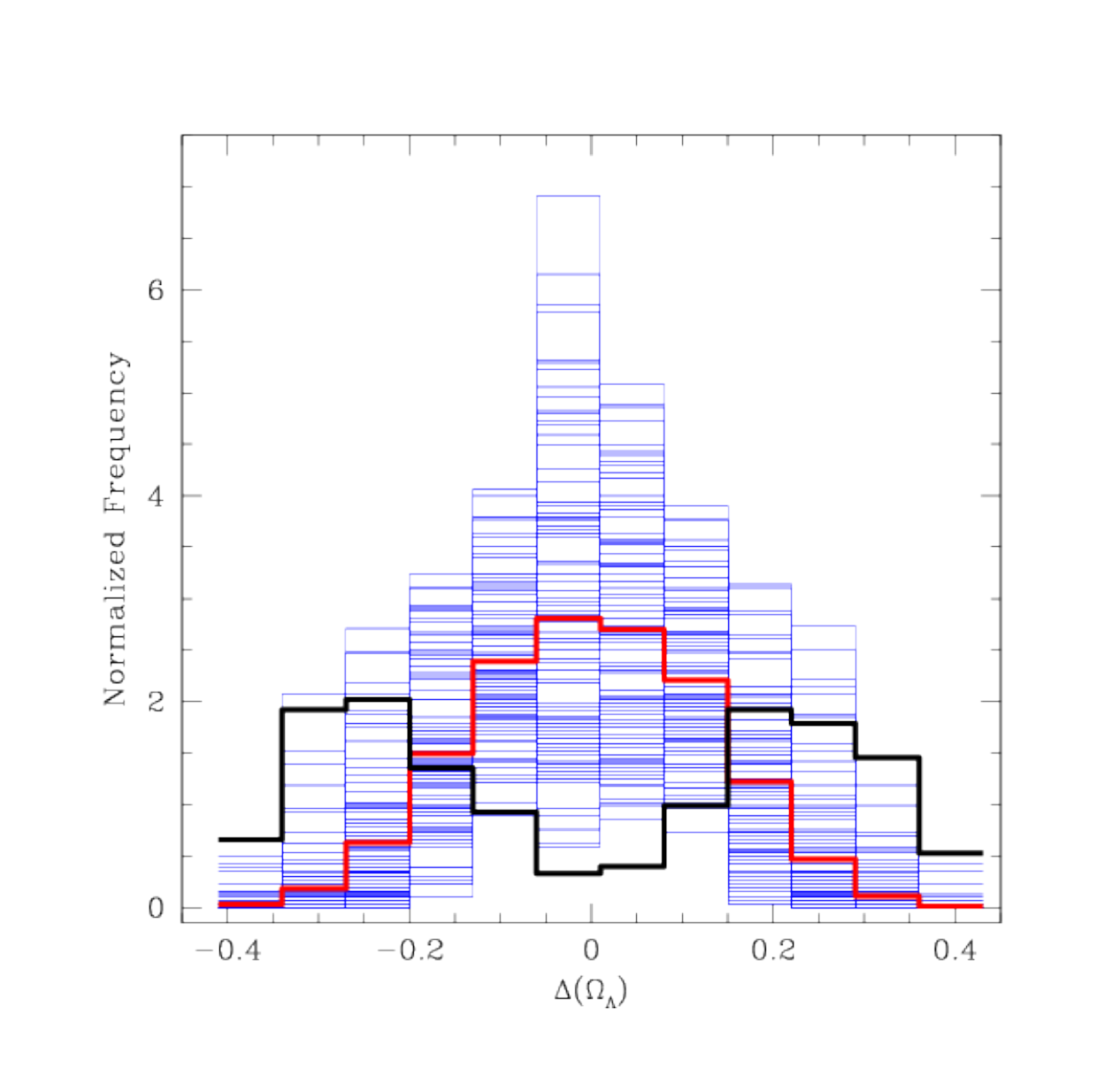}{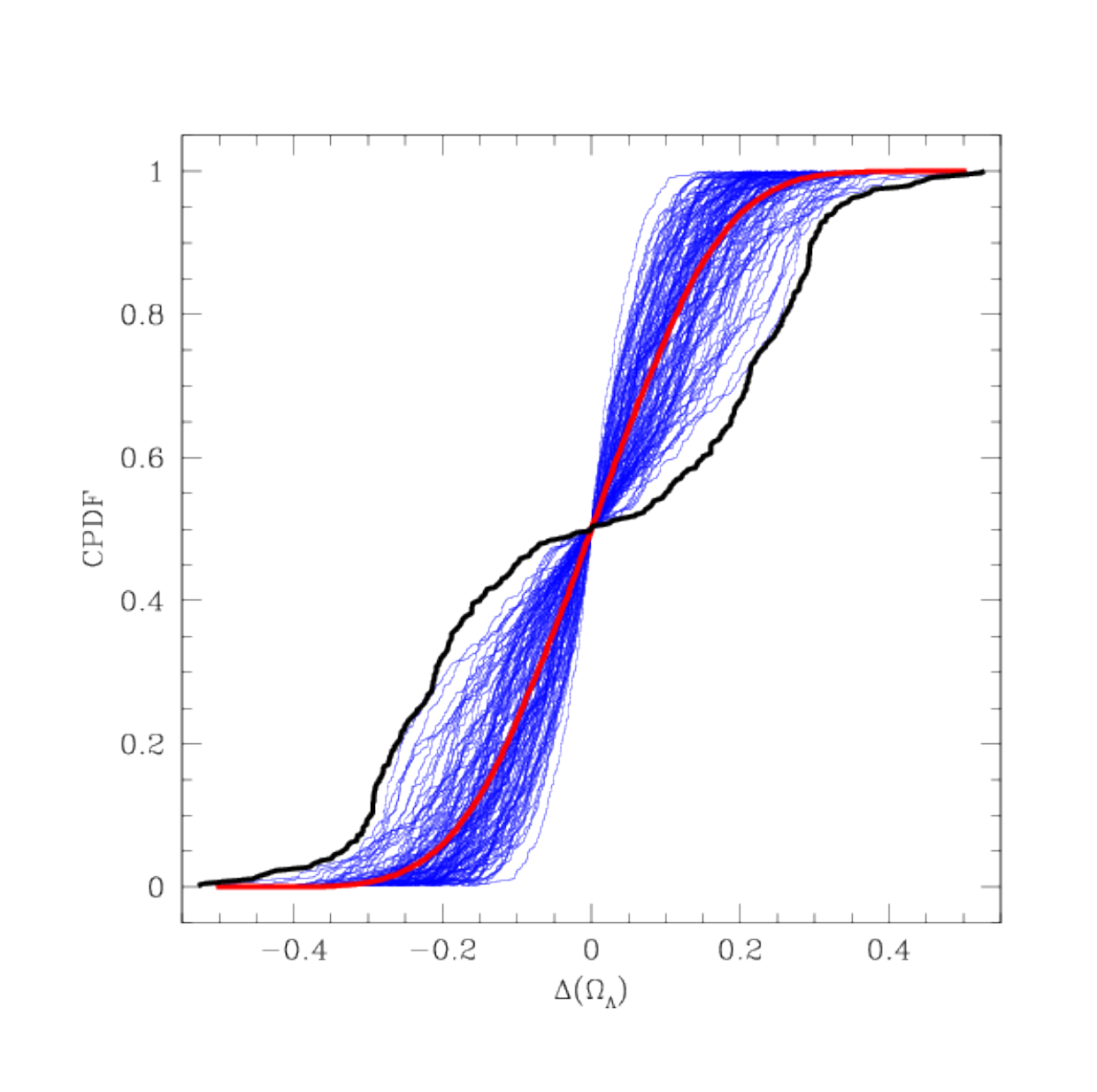}
\caption{
Same as Figure~\ref{fi:OML_stats} but now the histograms of the $\Omega_\Lambda$ differences in the pole-antipole directions are plotted (left panel).
As before, the right panel shows the same statistics through the Cumulative Probability density functions.
}
\label{fi:OML_diff_stats}
\end{figure}

An analysis of the Up-Down differences in the fake realizations of the sample, as shown in Figure~\ref{fi:OML_diff_stats}, provides a similar view.
Opposite to those of the real universe, whose bins are well populated towards the positive and negative sides, the typical histogram of the differences in the fake realizations is concentrated at zero, as expected for an isotropic acceleration.
Although there are a few flat, wide, and bimodal distributions, the histogram of the real universe is again the extreme only approached by in a handful of the fake cases.
It shows the largest Up-Down differences, the largest frequencies in the highest and lowest bins, and its CPDF is an outer envelope of those of the random SN distributions. 
The STDVs of the Up-Down difference histograms mirrors what is found for the fitted $\Omega_\Lambda$ histograms.
The real distribution of SNe in the sky carries the histogram with the largest STDEV (0.250) and 89 out of 100 random realizations have a STDEV smaller than 70\% of that value.

\subsection{Fitting Dipoles}

Fitting a dipole to the $\Omega_\Lambda$ signal brings some puzzling results.
Even though the largest observed value of $\Omega_\Lambda$ appears towards the region of the $\Omega_\Lambda$ Apex, and the largest contrast of any axis is between a nearby HEALPix pixel and its antipode, the signal that dominates the all sky fit comes from the large areas with high $\Omega_\Lambda$ towards the NGP and with low $\Omega_\Lambda$ towards the SGP. The amplitude of the variations along this axis is smaller than that in the $\Omega_\Lambda$ Apex axis, but the area covered is larger and the signal to noise is high. 

We fitted a dipole of the form
\begin{equation}
    \Omega_\Lambda = A \cos{\theta} + B ,
\label{eq:dipole}
\end{equation}
where $A$ is the amplitude of the dipolar signal, $B$ the monopole signal and $\theta$ the angle between an arbitrary point in the sky and the dipole direction, to the $\Omega_\Lambda$ map given in Figure~\ref{fi:mollweide_real}. We used the Downhill Simplex Method as coded in subroutine ``amoeba'' in   \protect\citet{1992nrfa.book.....P}. The fit was repeated 100 times, varying each individual $\Omega_\Lambda$ value at random, with a normal distribution with its corresponding, measured, dispersion.
The results are:
$A = 0.178 \pm 0.009$ and
$B = 0.586 \pm 0.004$,
and the direction of the dipole results RA$=219.8 \pm 2.8$, Dec$=19.0 \pm 2.7$.
%
The monopole term $B$ is consistent with the value of $\Omega_\Lambda$ measured for the whole Pantheon+ sample (see the Appendix) and the $A$ term implies a well detected variation of the total signal over the sky, with more than $\sim$60\% change from upper to lower extremes.
The fitted dipole direction is $\sim 26.6$ degrees from the NGP. The probability that this alignment is due to chance is smaller than 5.3\%.

If the previous fitting process is repeated with the signal of the $\Omega_\Lambda$ Apex region masked out, the results are
$A = 0.169 \pm 0.010$ and
$B = 0.571 \pm 0.004$,
with the dipole pointing towards RA$=212.4 \pm 3.2$, Dec$=26.4 \pm 2.6$.
The fitted dipole direction is now 18.0 degrees away from the NGP, with the probability of a chance alignment dropping to $\sim 2.5$\%.

We repeated the fit for all the realizations of the Pantheon+ sample with coordinates reshuffled, and analyzed the statistics of the fitted amplitudes. The results of this exercise lead to conclusions similar to those of comparing the statistics of fitted $\Omega_\Lambda$, or $\Delta (\Omega_\Lambda)$.   
Most of the Pantheon+ sample random reshufflings provide a weak dipolar signal and only a handful of them display a large asymmetry that can be fitted by a sizable dipole.

The anisotropic signal aligned  with the Milky Way axis is well fitted by a dipole and cannot be easily discarded as a chance event.
To trace a possible origin of this effect, we compared the distances of the Pantheon+ SN sample which are on the Northern Galactic hemisphere with those of the Southern ones.
We adopted a cosmology with $H_0=73 \, {\rm km \, s^{-1} Mpc^{-1}}$ as a ruler and found that, on average, each SN in the Northern Galactic hemisphere has a distance modulus ``excess'' of 0.016 magnitudes while those in the Southern Galactic hemisphere have an average excess distance modulus of only 0.010 mag.
Southern SNe appear, hence about six millimagnitudes closer to us on average and this is why the cosmological fits provide a lower $\Omega_\Lambda$ value.
Assuming a standard law for reddening, with $R_V \simeq 3$, the effect would imply that we are overestimating the color excess of the SNe on the Northern side of the Galaxy by approximately two millimagnitudes, on average, if compared with those on the Southern Galactic hemisphere. 

\begin{figure}[ht!]
\plottwo{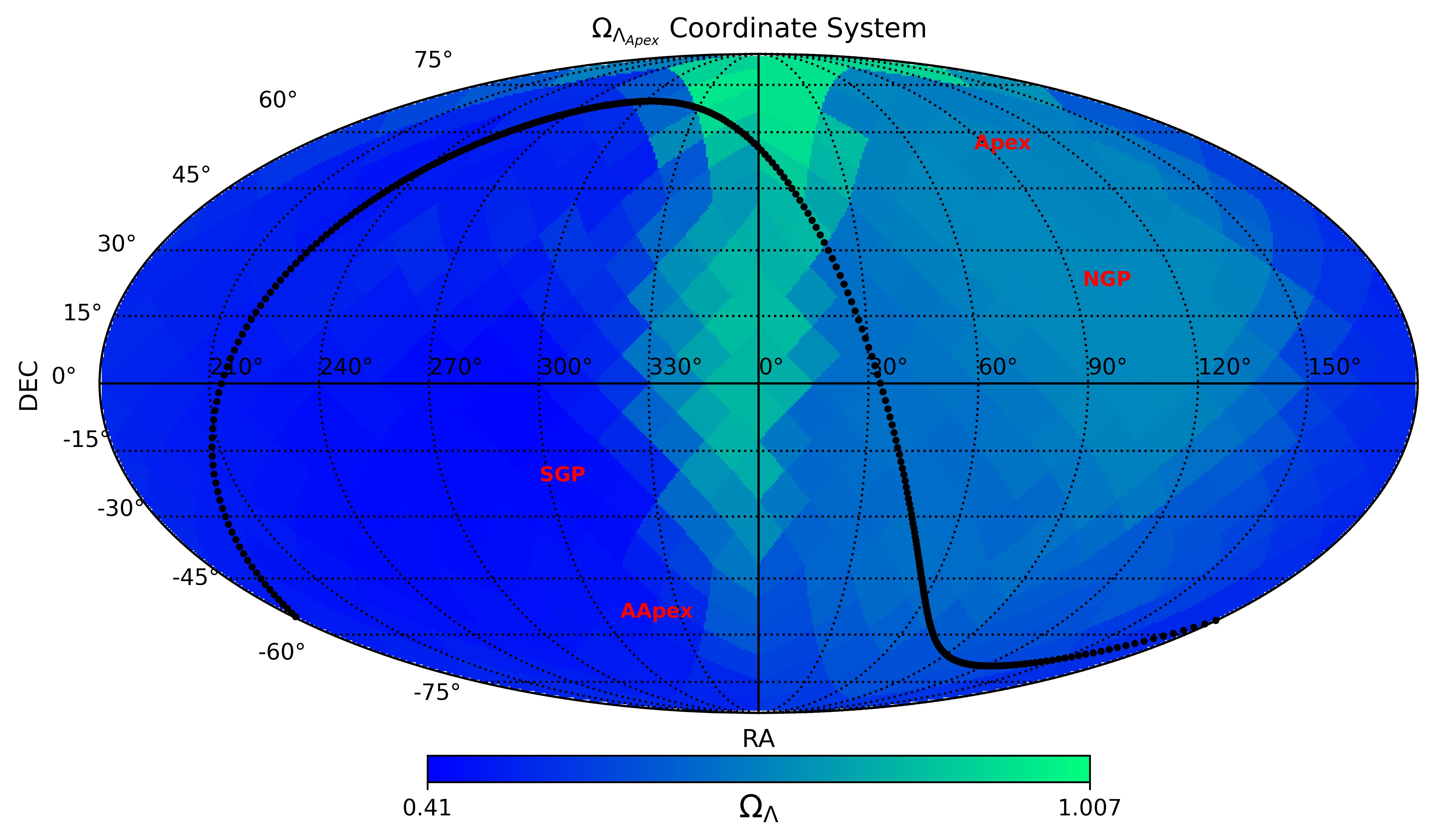}{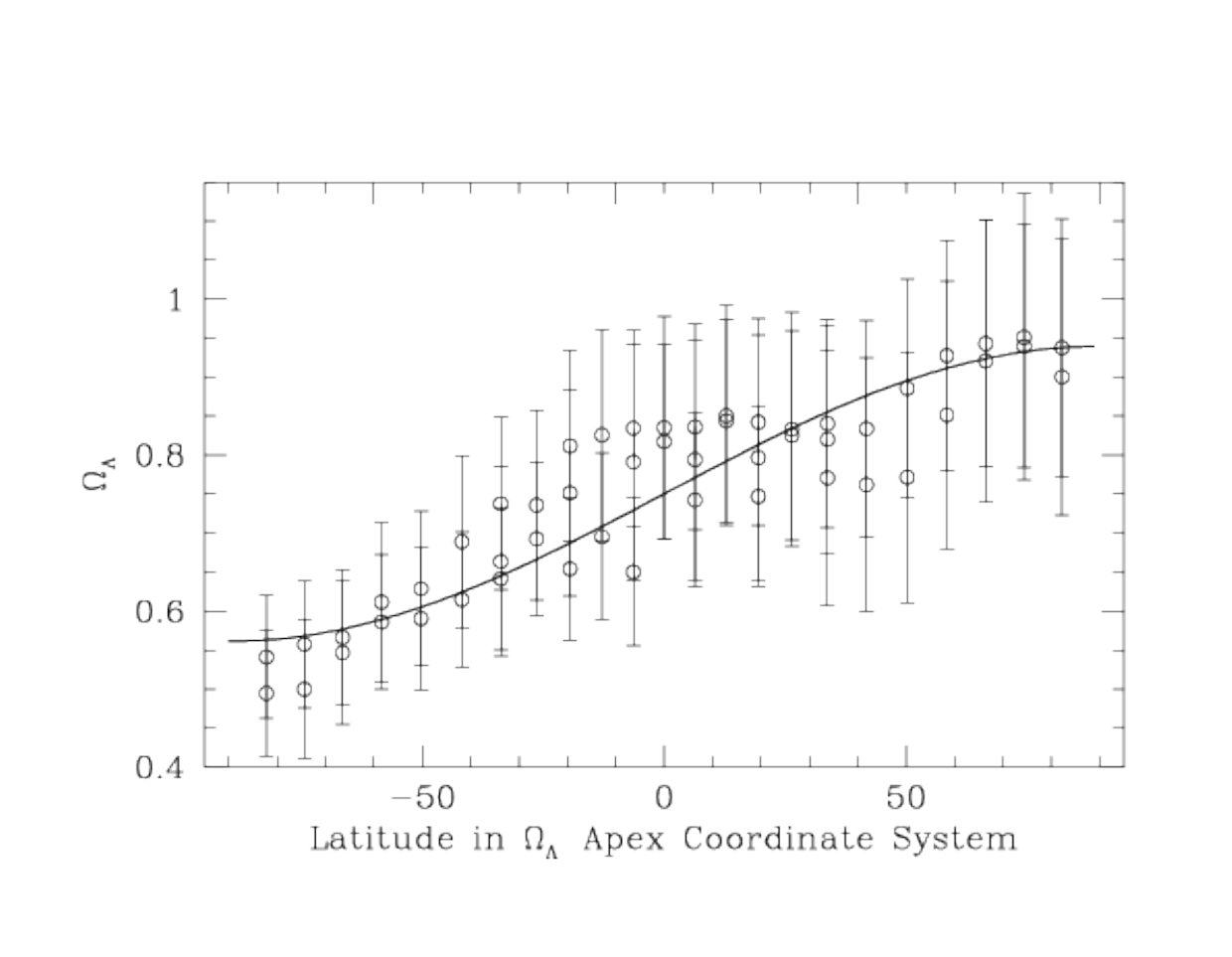}
\caption{The left panel shows the
$\Omega_\Lambda$ signal in Mollweide projection represented in the coordinate system where the $\Omega_\Lambda$ Apex direction (RA$\sim 217.5^\circ$ Dec$\sim -26.4^\circ$) is the upper pole.
The right panel shows the dipole fit in one dimension to the $\Omega_\Lambda$ signal of the HEALPix pixels inside the longitude zone $-20^\circ < l_{\rm D} < 20^\circ $ in the $\Omega_\Lambda$ Apex Coordinate System, including eight points within   $-45^\circ < l_{\rm D} < 45^\circ $ that are within 16 degrees of the poles.
The black solid line is the fit of equation \ref{eq:dipole} where $\theta$ equals the latitude like coordinate in the $\Omega_\Lambda$ Apex Coordinate System as seen in the left panel.
\label{fi:OML_coordinates}}
\end{figure}

Even though, when fitting a dipole to the $\Omega_\Lambda$ measurements over the whole sky, the large variation in the $\Omega_\Lambda$ Apex-Antiapex direction is overwhelmed by the signal in the NGP-SGP direction, by doing a one dimensional fit in a  judiciously chosen region it is possible to see that the former is also consistent with a dipole.
We transformed the Equatorial Celestial coordinates to a system where the Up pole is the $\Omega_\Lambda$ Apex (RA$ = 217.5^\circ$ Dec$ = -26.4^\circ$) to align the path from $\Omega_\Lambda$ Apex to $\Omega_\Lambda$ Antiapex with a meridian of the new system.
The result of this exercise is shown in the left panel of Figure~\ref{fi:OML_coordinates}.
This projection makes it easy to identify the region of the great circle passing through the $\Omega_\Lambda$ Apex and Antiapex poles that spans the maximum variation between the upper and lower values of $\Omega_\Lambda$.
We have left this region close to the origin (arbitrary) of the longitude-like variable ($l_{\rm D})$, and oriented $l_{\rm D} = 0$ towards the reader.
In the new projection, a one dimensional fit can be done by selecting the $\Omega_\Lambda$ values in the longitude-like zone where the strongest signal appears.
We took the 44 $\Omega_\Lambda$ values within the longitude zone $-20^\circ < l_{\rm D} < 20^\circ $ augmented by eight values in the longitude zone $-45^\circ < l_{\rm D} < 45^\circ $ which are closer than 16 degrees to the poles, an fitted a dipole-like form as given by equation~\ref{eq:dipole}.
In the new representation, $\theta$ is a latitude-like coordinate indicating the angle with the $\Omega_\Lambda$ Apex.
The result of this one dimensional fit is
$A = 0.189 \pm 0.032$,  $B = 0.750 \pm 0.019$,
and the fit is plotted in the right panel of Figure~\ref{fi:OML_coordinates}.
The monopole term is dominant, and $\sim 30$\% larger than the monopole fitted to the whole sphere, but a dipole-like signal that implies a peak-to-peak variation of $\sim 50$\% of the monopole signal is detected with S/N ratio of $\sim 6$.

We note that we have carried out a similar study selecting a different set of directions on the celestial sphere by using a Fibonacci Lattice   \protect\citep{Gonz_lez_2009} with 420 points.
The results of this experiment are qualitatively the same as those presented here with the most significant differences being the S/N ratio of the largest $\Delta \Omega_\Lambda$ difference, which results slightly smaller than 3, and the direction of the $\Omega_\Lambda$ Apex, which results RA$\sim 174^\circ$ Dec$\sim -32^\circ$ (i.e. $\sim 38^\circ$ from the axis found with the HEALPix pixelation presented here).

We finally note that our results are consistent with those of   \protect\citet{2012PhRvD..86h3517M}. They
also employed hemispheric comparison to analyze the pole--antipole differences in fitted cosmological parameters using the 557 SNe Union 2 sample   \protect\citep{2010ApJ...716..712A} and found a sizable anisotropy approximately aligned with the CMB dipole.
Our results, however, are opposite to those of   \protect\cite{2023ChPhC..47l5101T}, who perform a similar study using the same database but constraining the cosmological fit to reside on the line $\Omega_\Lambda + \Omega_{\rm M} = 1$. They conclude that the Pantheon+ sample is consistent with a large-scale isotropic universe, although they find evidence of a small amplitude dipole (with confidence level of 2-$\sigma$) by restricting the sample to SNe closer than $z = 0.1$.
A likely explanation to the different conclusions is that, by forcing the universe to be flat,   \protect\cite{2023ChPhC..47l5101T} are actually preventing the sample to display the anysotropies we describe here since they actually appear as a departure almost perpendicular to the line $\Omega_\Lambda + \Omega_{\rm M} = 1$ (see Figure \ref{fi:OM_high_low}).

\section{Conclusions} \label{se:final}

We have done an analysis of asymmetries of the cosmological constant $\Omega_\Lambda$ computed by fitting Hubble Diagrams on subsets of Type Ia SNe in the Pantheon+ sample that fall within hemispheres pointing towards 432 different directions in the sky.
We looked for large scale variations, in particular dipolar-like variations, superimposed on the expected signal for cosmic acceleration.
We have found that the monopole term is the dominant component, but also that $\Omega_\Lambda$ displays significant variations.
Surprisingly, these variations trace two approximately dipolar patterns.
One of them, the dominant, closely aligns with the NGP-SGP axis.
The second dipole, which carries the largest $\Omega_\Lambda$ values and largest axis contrast but appears covered by the first dipole and seems to span a smaller area on the sky, points towards RA$\sim 217.5^\circ$ Dec$\sim -26.4^\circ$, approximately 51 degrees from the CMB dipole Apex.
We called this direction the $\Omega_\Lambda$ Apex.

We tested the stability of these structures against variations of some of our assumptions and bootstrapping  and found them to be stable.
We developed an experiment consisting on randomly reshuffling the positions of the Pantheon+ SNe, recomputing the cosmology and comparing the results with those of the real universe (one hundred repetitions).
We found that the statistics of the real universe are usually the most extremes in terms of maximum value $\Omega_\Lambda$, pole-antipole differences and dispersion of the $\Omega_\Lambda$ histograms.
The CPDF of the real universe also stand out as extreme.
Typically, less than five out of 100 random shuffles of the Pantheon+ SN coordinates result in statistical parameters as extreme as those of the real universe.

The dominant dipolar signal picked up by an all-sky fit is the one aligned with the North-South Galactic axis.
This signal probably results from the combination of the natural concentration of SNe, especially those at large redhifts, at high galactic latitudes, with the techniques we are using to correct for Galactic extinction.
Puzzling as it is, it could be an indication of a difference between the Northern and Southern Galactic hemispheres (we would be overestimating the extinction in the Northern Galactic hemisphere in comparison with that of the Southern by $\sim 0.002$ magnitudes).
Fitting dipoles to the 100 samples made from random shuffles of the Pantheon+ SN coordinates shows that dipole signals as strong as the Galactic one in the real sample appear in a very small fraction of cases.

The other approximate dipole pointing $\sim 50$ degrees from CMB Apex implies a serious challenge to our interpretation of Cosmic Acceleration as a form of Dark Energy.
A dipolar signal in an observed global acceleration or deceleration is predicted for tilted observers in a FRW universe.
The $\Omega_\Lambda$ dipole traced by the SNe in the Pantheon+ sample, pointing approximately towards RA$ = 217.5^\circ$ Dec$ = -26.4^\circ$, suggests that we may be just that kind of tilted observers and, hence, the acceleration we measure is an apparent effect of the relativistic frame of reference transformation as predicted by   \protect\citet{2021EPJC...81..753T}.

If $\Omega_\Lambda$ is an apparent effect, a further test of the hypothesis proposed by   \protect\citet{2021EPJC...81..753T} is to calculate what should be the corresponding $w$ parameter in the EOS of the universe, as measured by tilted observers.
Since $w\simeq -1$ has been consistently measured for decades   \protect\citep[e.g.][]{2003ApJ...594....1T,2022ApJ...938..110B} this could serve to confirm or nullify this scenario.
Theory should as well be able to provide a quantitative estimate of the $\Omega_\Lambda$ dipole amplitude as a function of the bulk-flow velocities measured in the nearby universe.
This would strengthen the interplay between observation and theory and encourage, or discourage, further observations.

The result presented here as the difference { $\Delta \Omega_\Lambda = 0.48 \pm 0.17$ in the $\Omega_\Lambda$ Apex--Antiapex axis implies a 2.8-$\sigma$ tension regarding the hypothesis that $\Omega_\Lambda$ is a constant} and it calls for our earnest attention to the issue of anysotropies in the Cosmological Constant.
Ultimately, independently of the theory we use to interpret the observations, the confirmation of these puzzling results would come from collections of SN samples like the Pantheon+, but concentrated in small regions of the sky conveniently located.
Building these samples around the $\Omega_\Lambda$ Apex and Antiapex directions appears as the first priority, since these regions provide the largest difference in fitted $\Omega_\Lambda$.

\appendix
\label{se:appendix}
{\hfill EFFECT OF USING, OR NOT, THE COVARIANCES AND SYSTEMATIC UNCERTAINTIES\hfill}

\begin{figure}[ht!]
\plotone{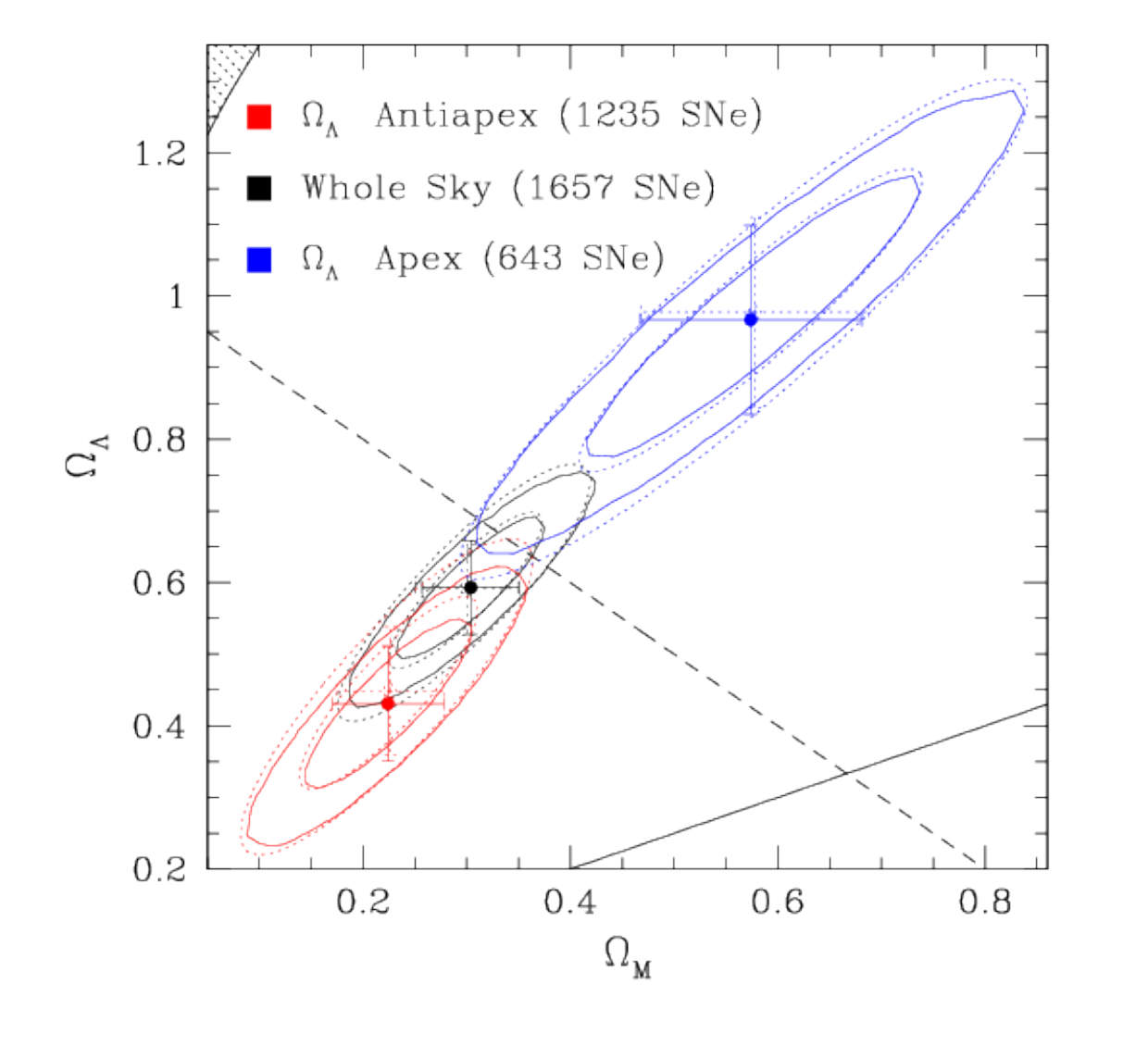}
\caption{
Confidence contours at the 68.3 \% and 95.4 \% levels for the $\Omega_\Lambda$ and $\Omega_{\rm M}$ parameters for the whole Patheon+ SN sample (black lines), the subsample in the $\Omega_\Lambda$ Apex direction (blue lines), and the $\Omega_\Lambda$ Antiapex direction (red lines).
The solid lines correspond to the computing using the full $C_{\rm stat}$ covariance matrix, and the short dashed lines to using only the diagonal elements.
The solid and dashed diagonal straight lines and the small shaded region at the top-left corner have  the same meaning as in Figure~\ref{fi:OM_high_low}.
}
\label{fi:compApexAapex}
\end{figure}

\bigskip
To build the hemispherical analysis we took two decisions regarding the data which needed additional justification.
Firstly, since we were to generate fake distributions of the SNe scrambling their positions in the sky we decided to use only the statistical uncertainties and not the systematic ones, since the latter are more closely related to where the objects are located in the sky.
Secondly, we decided to carry on just the diagonal of the covariance matrix, and not to use the covariances between  events, to reduce the computing load.
The main justification to do so is that the effect of the covariances, while relevant in the quest to have a precise measurement of the cosmological parameters, is marginal if compared with the large variations resulting from splitting the SN sample in hemispheres.
The referee essentially asked us to show the effect of these assumptions, so we produced this appendix, Figure~\ref{fi:compApexAapex}, and the right panel of Figure~\ref{fi:OM_high_low}.


In Figure~\ref{fi:compApexAapex} we display both the confidence contours computed using just the diagonal of the STAT\_ONLY covariance matrix {($C_{\rm stat}$)} of the Pantheon+ SN sample provided by   \protect\citet{2022ApJ...938..110B} (short dashed lines) and those resulting from using the whole {$C_{\rm stat}$ matrix} (solid lines).
{In the last case, the confidence contours are obtained from the posterior probability distribution $P=p\mathcal{L}$, where $\mathcal{L}$ is the likelihood function given by Equation~15 of   \protect\citet{2022ApJ...938..110B} (using $C_{\rm stat}$ as covariance matrix) and $p$ is the prior function of the parameters. We adopt a conservative uniform prior, which is equal to a positive constant for $\Omega_{\rm M}$ ($\Omega_\Lambda$) between 0.0 and 1.0 (1.6) and equal to zero otherwise. We sample $P$ by means of a Markov Chain Monte Carlo (MCMC) process using the Python package \texttt{emcee}   \protect\citep{2013PASP..125..306F}, running for 100,000 steps. Then, we discard the initial 150 steps and thin by half the autocorrelation time (25 steps).}
Black lines correspond to the whole Pantheon+ SN sample, blue and red lines correspond to the hemispheres with poles that we called $\Omega_\Lambda$ Apex and Antiapex, respectively (see \S \ref{ss:results}).
We compared our solid black contours with those of   \protect\citet{2022ApJ...938..110B} (the unfilled dashed contours in their Figure~8) and found them to coincide within the width of the lines.
In columns 4-7 of Table~\ref{ta:compApexAapex} we provide the fitted cosmological parameters corresponding to each sample and technique used.
It is clear from Figure~\ref{fi:compApexAapex} and the table that the effect on the fitted cosmological parameters of using, or not using, the whole covariance matrix is very minor in comparison with the effect of selecting samples in different hemispheres.

To study the difference that the systematic uncertainties make in our analysis we repeated the MCMC process but now using the complete statistical plus systematic covariance matrix, {$C_{\rm stat+sys}$}   \protect\citep[eq. 8 in][]{2022ApJ...938..110B}.
The results are shown in the right panel of Figure~\ref{fi:OM_high_low} and columns 2 and 3 of Table~\ref{ta:compApexAapex}.
Although the differences between cosmological parameters fitted with the different techniques is always smaller than the uncertainties, they have some bearing on the contrast between the $\Omega_\Lambda$ fitted in the Apex and Antiapex.
$\Delta \Omega_\Lambda$ is a 3.3-$\sigma$ result when using the diagonal of $C_{\rm stat}$, a 3.5-$\sigma$ result when using the whole $C_{\rm stat}$ matrix, and a 2.8-$\sigma$ result when using the complete $C_{\rm stat+sys}$ matrix.


\begin{table}[h]
\centering
\begin{tabular}{|c|c|c|c|c|c|c|c|}
\toprule
Sample & $\Omega_{\rm M}$ & $\Omega_\Lambda$ & $\Omega_{\rm M}$ & $\Omega_\Lambda$ & $\Omega_{\rm M}$ & $\Omega_\Lambda$ & N SNe\\

       & \multicolumn{2}{c}{($C_{\rm stat+sys}$ whole matrix)} &
        \multicolumn{2}{c}{($C_{\rm stat}$ whole matrix)} &  \multicolumn{2}{c}{($C_{\rm stat}$ diagonal)}  & \\
\hline
Whole sky & 0.305$\pm$0.054 & 0.628$\pm$0.080 & 0.304$\pm$0.047 & 0.593$\pm$0.066 & 0.300$\pm$0.050 & 0.592$\pm$0.067 & 1657 \\
\hline
$\Omega_\Lambda$ Apex & 0.560$\pm$0.111 & 0.979$\pm$0.142 & 0.574$\pm$0.107 & 0.967$\pm$0.132 & 0.578$\pm$0.110 & 0.977$\pm$0.132 & 643 \\
\hline
$\Omega_\Lambda$ Antiapex & 0.242$\pm$0.061 & 0.502$\pm$0.096 & 0.224$\pm$0.054 & 0.431$\pm$0.080 & 0.226$\pm$0.052 & 0.449$\pm$0.089 &1235 \\
\bottomrule
\end{tabular}
\caption{Cosmological parameters fitted using the complete $C_{\rm stat+sys}$ covariance matrix, those fitted using only $C_{\rm stat}$ and those fitted using only the diagonal of $C_{\rm stat}$.}
\label{ta:compApexAapex}
\end{table}

\begin{acknowledgments}
We acknowledge the work of the anonymous referee who helped us to produce an improved version the paper.
The work of AC, OR and AOM has been supported by ANID Millennium Institute of Astrophysics (MAS) under grant ICN12\_009.
The Geryon
cluster at the Centro de Astro-Ingenieria UC was extensively used for the 
calculations performed in this paper. BASAL CATA PFB-06, the Anillo ACT-86, 
FONDEQUIP AIC-57, and QUIMAL 130008 provided funding for several improvements 
to the Geryon cluster.
This research has made use of NASA’s Astrophysics Data System.

\end{acknowledgments}

\bibliography{DOSC}{}
\bibliographystyle{aasjournal}
\end{document}